# Are Large Language Models a Threat to Programming Platforms? An Exploratory Study

Md Mustakim Billah
University of Saskatchewan
Canada
mustakim.billah@usask.ca

Palash Ranjan Roy
University of Saskatchewan
Canada
palash.roy@usask.ca

Zadia Codabux
University of Saskatchewan
Canada
zadiacodabux@ieee.org

Banani Roy
University of Saskatchewan
Canada
banani.roy@usask.ca

## Abstract

**Background:** Competitive programming platforms such as LeetCode, Codeforces, and HackerRank provide challenges to evaluate programming skills. Technical recruiters frequently utilize these platforms as a criterion for screening resumes. With the recent advent of advanced Large Language Models (LLMs) like ChatGPT, Gemini, and Meta AI, there is a need to assess their problem-solving ability on the programming platforms. **Aims:** This study aims to assess LLMs' capability to solve diverse programming challenges across programming platforms with varying difficulty levels, providing insights into their performance in real-time and offline scenarios, comparing them to human programmers, and identifying potential threats to established norms in programming platforms. **Method:** This study utilized 98 problems from LeetCode and 126 from Codeforces, covering 15 categories and varying difficulty levels. Then, we participated in nine online contests from Codeforces and LeetCode. Finally, two certification tests were attempted on HackerRank to gain insights into LLMs' real-time performance. Prompts were used to guide LLMs in solving problems, and iterative feedback mechanisms were employed. We also tried to find any possible correlation among the LLMs in different scenarios. **Results:** LLMs generally achieved higher success rates on LeetCode (e.g., ChatGPT at 71.43%) but faced challenges on Codeforces. While excelling in HackerRank certifications, they struggled in virtual contests, especially on Codeforces. Despite diverse performance trends, ChatGPT consistently performed well across categories, yet all LLMs struggled with harder problems and lower acceptance rates. In LeetCode archive problems, LLMs generally outperformed users in time efficiency and memory usage but exhibited moderate performance in live contests, particularly in harder Codeforces contests compared to humans. **Conclusions:** While not necessarily a threat, the performance of LLMs on programming platforms is indeed a cause for concern. With the prospect of more efficient models emerging in the future, programming platforms need to address this issue promptly.

## 1 Introduction

Programming platforms, also known as online judges [49], such as Codeforces[1], LeetCode[2], and HackerRank[3], are widely used as tools for developing and demonstrating programming skills. Success on these platforms often leads to recognition within the programming community, as programmers earn scores based on their performance [3]. Moreover, technical recruiters, including top companies, frequently use competitive programming as a benchmark to evaluate candidates' suitability for software engineering positions [15, 28]. Given the time-consuming nature of manual resume screening, recruiters rely on automatic techniques [18, 36] that prioritize candidates' online programming activities to expedite the hiring process. In academia, high performance and contributions on these platforms can significantly influence author or study rankings [38], showcasing expertise and innovation in the field.

In recent years, Large Language Models (LLMs) have demonstrated promising performance in code generation [2, 10, 23, 24, 37, 46]. Additionally, companies such as OpenAI, Google, and Meta have released powerful conversational LLMs, including ChatGPT[4],

---

[1]https://codeforces.com
[2]https://leetcode.com
[3]https://www.hackerrank.com
[4]https://chat.openai.com

Gemini[5], and Meta AI[6], which introduce a new potential for dishonesty in programming challenges on online platforms. This could potentially mislead recruiters seeking skilled employees for their companies in resume screening. Misjudgment in recruitment can significantly hamper the operations and success of a software company. This can lead to a range of negative consequences, including decreased productivity, increased turnover [35], and a decline in overall company morale.

Several studies have explored the problem-solving capabilities of LLMs on programming platforms [1, 7, 29, 40]. However, these studies often concentrate on assessing a single platform, typically LeetCode, or a single LLM, frequently ChatGPT, and they tend to focus on specific programming challenges. We contend that these findings may not generalize to other scenarios, as the capabilities of the same LLM can vary significantly across different platforms. Similarly, platforms may yield different results when tested with different LLMs. Therefore, conducting comprehensive research that encompasses a diverse range of programming challenges across multiple online platforms and evaluating the impact of LLMs on these platforms will be beneficial for both platform maintainers and technical recruiters.

In this study, we investigated three types of programming challenges across multiple programming platforms using three conversational LLMs – ChatGPT, Gemini [43], and Meta AI [45]. Initially, we compiled a dataset consisting of 224 programming problems spanning 15 categories, sourced from online judge archives. This dataset comprised 98 problems from LeetCode and 126 from Codeforces. Subsequently, we engaged in nine virtual contests, encompassing a total of 49 problems on Codeforces and LeetCode, to replicate real-time programming conditions. Additionally, we completed two certification tests on HackerRank, each comprising four problems in real-time settings.

To the best of our knowledge, this is the first attempt to explore the impact of different LLMs on programming platforms by analyzing varied programming challenges across different platforms. Our key contributions include the following:

(1) Evaluation of LLMs' problem-solving abilities across a range of programming challenges from judge archives and online contests.
(2) Analysis of variations in LLMs' performance across different problem dimensions, such as category, difficulty level, acceptance rate, and programming language.
(3) Comparison of LLMs' problem-solving performance with that of human programmers, with implications for online judging systems and recruiters.
(4) Comparative analysis of LLMs' performance, examining significant differences in problem-solving abilities across various LLMs.
(5) Provision of a comprehensive replication package[7].

The rest of the paper is organized as follows. Section 2 provides background information for our study. Section 3 describes related works. Section 4 outlines the methodology. Sections 5 and 6 present, analyze and discuss our findings. Section 7 lists the implications of our study. Section 8 states the limitations of our study. Finally, Section 9 concludes the study and suggests future research directions.

## 2 Background

### 2.1 LLMs for Code Generation

LLMs have seen rapid development, with models like CodeGen [32], StarCoder [21], WizardCoder [25], CodeT5 [48], CodeT5+ [47], and Incoder [12] specifically designed for code generation. However, these models lack plug-and-play features, requiring a certain level of expertise for downloading and utilization. Conversely, some LLMs, such as GitHub Copilot[8] and Amazon CodeWhisperer[9], are seamlessly integrated into developers' IDEs as code assistants.

In this study, we focus on conversational LLMs only, which are easily accessible without much expertise. By conversational LLMs, we refer to online browser versions like ChatGPT, Gemini, and Meta AI, which enable users to interact with them through prompts and foster conversational interactions.

### 2.2 Reasoning With LLMs

Solving complex problems on programming platforms necessitates human reasoning and logical analysis. This involves devising logical pathways while considering various scenarios and learning from error feedback provided by the platforms. Wei et al. [50] demonstrated that LLMs can effectively address challenging reasoning tasks by systematically guiding their logic through chain-of-thought prompting. This approach has notably enhanced mathematical problem-solving skills at the high school level.

In our problem-solving approach (Section 4.4), we adopt a similar chain-of-thought prompting strategy with LLMs, providing them with error feedback to facilitate improved reasoning.

### 2.3 Online Judge

Programming platforms like Codeforces, LeetCode, and HackerRank, also known as online judges, assess submitted programming solutions. These platforms host various problem types, each categorized under a specific problem type and requiring different skill sets for effective resolution. For instance, the "Greedy" category involves developing algorithms based on locally optimal choices at each stage.

The online judge-checking system thoroughly evaluates submitted solutions to verify correctness and adherence to problem requirements. The system issues a verdict, indicating acceptance or rejection based on specific criteria. Verdicts typically classify errors into four types: Time Limit Exceeded and Memory Limit Exceeded occur when solutions exceed time or memory constraints, Compilation Error results from code compilation issues, and Runtime Error arises during code execution.

## 3 Related Work

### 3.1 ChatGPT and LeetCode Based Studies

Previous studies primarily concentrated on using ChatGPT to solve problems from LeetCode and assessing its performance. Sakib et al. [40] tried to solve 128 problems of LeetCode. They selected ten different types of problems and attempted to solve them twice. They reported that ChatGPT solved 71.9% of their problems and could improve solutions for the unsolved problems up to 36.7% in the second attempt. Nascimento et al. [29] tried one LeetCode contest and compared their results with 42 participants. They concluded that ChatGPT improves the performance of easy and medium-level problems compared to novice contest programmers. Ekedahl et al. [9] also tried 90 LeetCode problems of ten types by GPT-4 version and tried for a maximum of three times if a solution failed. They showed that ChatGPT excels in simpler and medium-level programming but struggles with harder challenges, showing declining accuracy.

---

[5]https://gemini.google.com/app
[6]https://www.meta.ai/
[7]https://github.com/srlabUsask/LLM_Threat_Programming_Platforms

[8]https://github.com/features/copilot
[9]https://aws.amazon.com/fr/codewhisperer/

## 3.2 Comparative Studies

Several comparative studies [1, 4, 41, 42] have examined ChatGPT, Bard (now known as Gemini), and Llama across various dimensions. Ahmed et al. [1] specifically compared ChatGPT and Bard, identifying a significant shared limitation termed "Artificial Hallucinations." However, they observed that GPT-4 produces more accurate solutions with fewer hallucinations compared to Bard.

Nikolaidis et al. [33] focused on 50 LeetCode problems, concluding that ChatGPT and Copilot excel in providing Java and Python solutions but exhibit decreased performance with C language. Similarly, Hans et al. [14] discovered, through experiments with 80 medium and 60 easy LeetCode problems, that ChatGPT significantly outperforms Bard.

Recently, Coignion et al. [7] evaluated 18 LLMs for code generation efficiency using a LeetCode dataset, revealing comparable performance with human-crafted solutions and indicating potential for future optimizations. research conducted by Nascimento et al. [30] and Idrisov et al. [17] has compared LLMs with human performance, highlighting certain shortcomings of LLMs.

## 3.3 Generative Models

Li et al. [22] introduced AlphaCode, a deep learning-based code generation system, which achieved an average ranking within the top 54.3% in simulated evaluations during recent Codeforces programming competitions. OpenAI's Codex [11] and Github's Copilot [8] are also capable code generation models. Chen et al. [6] evaluated Codex and found that it has a strong performance for easy interview problems. Nguyen et al. [31] evaluated the suggested codes of GitHub Copilot with LeetCode and found that it achieved at most 57% correctness score.

In a case study, Lertbanjongngam et al. [20] found that AlphaCode's performance is comparable to or sometimes worse than humans regarding execution time and memory usage. They also noted that AlphaCode often utilizes too many nested loops and unnecessary variable declarations for high-difficulty problems. Furthermore, they noted that AlphaCode is capable of generating entire programs from lengthy natural language descriptions, distinguishing it from Codex and GitHub Copilot.

***Summary:*** Unlike previous research, our study seeks to offer comprehensive insights to both programming platform maintainers and technical recruiters. We achieve this by analyzing different online judges across diverse problem categories. Our investigation spans various challenges, such as solving archived problems, participating in real-time virtual contests, and engaging in certification programs. Our sole aim is to evaluate the impact of LLMs on well-known programming platforms.

## 4 Methodology

This section outlines the study's approach, detailing its aims, research inquiries, data collection, and analysis procedures.

### 4.1 Goal

The goal of the study is described using the Goal-Question-Metric technique [5] as follows:

> **Purpose:** To evaluate
> **Issue:** The impact of LLMs
> **Object:** On programming platforms
> **Viewpoint:** From the perspective of practitioners.

### 4.2 Research Questions

Based on our goal, we derive the following Research Questions (RQs):

**RQ$_1$: How successfully can LLMs solve a diverse set of programming challenges?**

To evaluate LLMs' problem-solving capabilities, we aim to assess their performance across various types of programming challenges. This research question will provide insights into how LLMs perform when tackling problems of different complexities in both offline and online environments. A high success rate could indicate potential vulnerabilities in online judges, as people might exploit these judges using LLMs. Conversely, a lower success rate would demonstrate that online judges remain robust against LLMs.

**RQ$_2$: How does the performance of LLMs vary across different dimensions?**

We hypothesize that LLMs will exhibit varied performance based on problem category, difficulty level, acceptance rate, and programming language. Additionally, LLMs may differ in their performance from one another.

**RQ$_3$: How does the problem-solving performance of LLMs compare to that of human programmers?**

Investigating this question will offer insights into how the problem-solving abilities of each LLM stack up against those of human programmers, thereby highlighting potential implications for online judging systems.

### 4.3 Data Collection

*4.3.1 Problem Category Selection.* As of April 26, 2024, LeetCode's collection comprises 3125 problems spread across 71 distinct categories, while the Codeforces archive contains 9460 problems categorized into 37 distinct types. For this study, we have selected 15 popular[10] categories that are present on both platforms. These categories include String (S), Two Pointers (TP), Math (M), Greedy (G), Binary Search (BS), Combinatorics (C), Depth-First Search (DFS), Divide and Conquer (DC), Dynamic Programming (DP), Matrix (MA), Sorting (SO), Number Theory (N), Shortest Path (SP), Probability and Statistics (PS), and Tree (T).

---

[10]https://en.wikipedia.org/wiki/Competitive_programming#Overview

*4.3.2 Problems from Judge Archives.* In LeetCode, problems are categorized into three difficulty levels: Easy, Medium, and Hard. Each problem also has an associated acceptance rate, indicating the percentage of times the problem was solved compared to the total submissions by users. Our dataset was structured based on these difficulty levels and acceptance rates, following the approach outlined by Sakib et al. [40]. They defined three acceptance range tiers: High (>70%), Medium (>=30% and <=70%), and Low (<30%). The goal was to include at least one problem from each of the 15 categories, covering different difficulty levels and acceptance ranges. By following this methodology, we compiled a set of 98 problems, as depicted in TABLE 1.

Codeforces does not organize problems by difficulty levels or display acceptance rates. Instead, it assigns ratings[11] to indicate complexity and solution frequency. In our study, we classified ratings into Easy (800-1200), Medium (1201-1700), and Hard (exceeding 1700). To identify problems with High, Medium, and Low acceptance rates, we sorted them based on successful solutions and chose one problem per rate category. This process resulted in a total of 126 problems (TABLE 1) selected for our study.

*4.3.3 Online Contests.* In addition to the problems obtained from judge archives, we actively participated in the latest online contests, particularly on platforms like Codeforces and LeetCode. Codeforces hosts various types of contests, ranging from Division 1 (the most challenging) to Division 4 (less challenging), along with unrated Educational contests. We selected five recent contests from Codeforces, representing each contest type, and competed in them virtually. Virtual contests provide a real-time competition experience against participants worldwide within a timed setting.

Moreover, we engaged in LeetCode's weekly and biweekly contests, totaling four recent contests, with two for each contest type, through virtual participation. This effort resulted in our involvement in a total of nine online contests for our study.

*4.3.4 Certification Tests.* We opted for HackerRank for certification tests, considering its well-established reputation as a reliable platform for assessing problem-solving skills through certifications. These tests simulate real-time contests, presenting participants with a series of problems to solve within specified time constraints. However, unlike our approach with Codeforces and LeetCode, where we selected problems from their archives, we did not directly solve any problems from HackerRank's archive for this study. Instead, we undertook two problem-solving-based certification tests on the platform. Further details about these tests are provided in TABLE 2.

## 4.4 Solving Problems

We utilized open-source, free conversational language models such as ChatGPT-3.5, Gemini 1.0 Pro, and Meta AI for our study. To conduct our experiment using three different conversational LLMs, we require a standardized prompt structure. A prompt serves as a set of instructions given to an LLM, programming it by customization and/or enhancing its capabilities [51]. Following the approach outlined in a recent study [16], we closely followed the guidelines[12] provided by OpenAI, as detailed in their documentation on prompt engineering to optimize the prompts to get better results from the LLMs.

Each problem on these platforms comes with its own constraints regarding memory usage and execution time. These constraints establish the parameters within which a solution must operate. This means that even if a solution is technically correct, it will not be accepted if it exceeds the specified time limit or consumes more memory than allowed. The prompts for solving these problems had to be formatted differently based on the specific requirements of each online judge. For instance, on Codeforces, the problem descriptions typically start by outlining the time and memory limit constraints, followed by a detailed problem description. Solutions on this platform are not required to be written within predefined functions; instead, a complete solution is expected from start to finish. An example of such a prompt is illustrated in the upper part of Figure 1.

Conversely, platforms like LeetCode and HackerRank necessitate solutions to be written within specific predefined functions provided by the platform. Additionally, these platforms incorporate time and memory limit constraints within the problem description itself. This difference in approach results in distinct prompt designs, as depicted in the lower part of Figure 1 compared to Codeforces. We chose C++ as the primary language for our study because competitive programmers commonly use it for problem-solving [16].

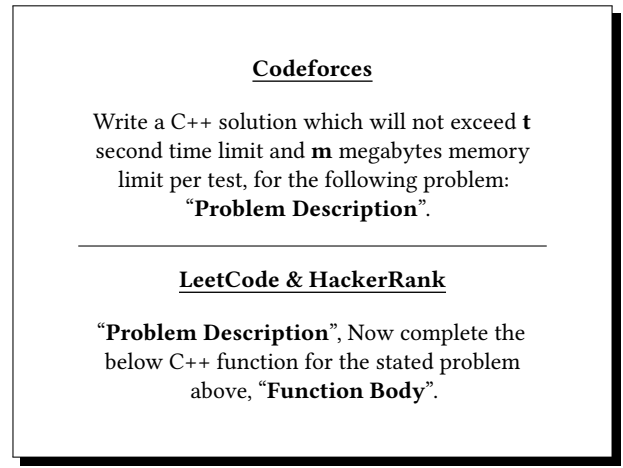

Figure 1: Prompt used for different platforms

After receiving a solution from an LLM, we submitted it to the relevant judges to determine whether it would be accepted or rejected. Rejection could occur for various reasons such as Wrong Answer, Time Limit Exceeded, Memory Limit Exceeded, Compilation Error, or Runtime Error. If rejected, we provided feedback to the LLMs, as outlined in Madaan et al. [26], which included specific error details to help them address the issues. Following this feedback, the LLMs then generated an alternative solution, incorporating the insights gained from our guidance.

The workflow of our methodology is illustrated in Figure 2. We iteratively employed few-shot prompting to tackle a problem. If an LLM consistently produced incorrect solutions for k consecutive attempts, we stopped the process. In our experiment, we set the value of k to 5, indicating that we attempted each problem five times before discontinuing the process.

---
[11]https://en.wikipedia.org/wiki/Codeforces#Rating_system
[12]https://platform.openai.com/docs/guides/prompt-engineering

Table 1: Cross Platform Dataset Across Problem Domains and Difficulty Levels

| Difficulty | Acceptance | LeetCode \| Codeforces | | | | | | | | | | | | | | |
|---|---|---|---|---|---|---|---|---|---|---|---|---|---|---|---|---|
| | | S | TP | M | G | BS | C | DFS | DC | DP | MA | SO | N | SP | PS | T |
| Easy | High | 1\|1 | 1\|1 | 1\|1 | 1\|1 | 1\|1 | 1\|1 | 1\|1 | 1\|1 | 1\|1 | 1\|1 | 1\|1 | 1\|1 | 0\|1 | 0\|1 | 1\|1 |
| | Medium | 1\|1 | 1\|1 | 1\|1 | 1\|1 | 1\|1 | 0\|1 | 1\|1 | 1\|0 | 1\|1 | 1\|0 | 1\|1 | 1\|1 | 0\|0 | 0\|1 | 1\|1 |
| | Low | 1\|1 | 0\|1 | 1\|1 | 1\|1 | 0\|1 | 0\|1 | 0\|1 | 0\|0 | 0\|1 | 0\|0 | 0\|1 | 0\|1 | 0\|1 | 0\|1 | 0\|0 |
| Medium | High | 1\|1 | 1\|1 | 1\|1 | 1\|1 | 0\|1 | 1\|1 | 1\|1 | 1\|1 | 1\|1 | 1\|1 | 1\|1 | 0\|0 | 0\|0 | 0\|0 | 1\|1 |
| | Medium | 1\|1 | 1\|1 | 1\|1 | 1\|1 | 1\|1 | 1\|1 | 1\|1 | 1\|1 | 1\|1 | 1\|1 | 1\|1 | 1\|1 | 1\|1 | 1\|1 | 1\|1 |
| | Low | 1\|1 | 1\|1 | 1\|1 | 1\|1 | 1\|1 | 0\|1 | 1\|1 | 0\|1 | 1\|1 | 0\|1 | 1\|1 | 1\|1 | 1\|1 | 0\|1 | 0\|1 |
| Hard | High | 1\|1 | 0\|1 | 0\|1 | 1\|1 | 1\|1 | 0\|1 | 1\|1 | 0\|1 | 0\|1 | 1\|1 | 1\|1 | 0\|1 | 1\|1 | 0\|1 | 0\|1 |
| | Medium | 1\|1 | 1\|1 | 1\|1 | 1\|1 | 1\|1 | 1\|1 | 1\|1 | 1\|1 | 1\|1 | 1\|1 | 1\|1 | 1\|1 | 1\|1 | 1\|1 | 1\|1 |
| | Low | 1\|1 | 1\|1 | 1\|1 | 1\|1 | 1\|1 | 1\|1 | 0\|1 | 1\|1 | 1\|1 | 1\|1 | 1\|1 | 1\|1 | 1\|1 | 0\|1 | 1\|1 |
| Total Problems | | 9\|9 | 7\|9 | 8\|9 | 9\|9 | 7\|9 | 5\|9 | 7\|9 | 6\|7 | 7\|9 | 7\|7 | 8\|9 | 6\|8 | 5\|7 | 2\|8 | 5\|8 |

Table 2: HackerRank Data

| Certification Name | Duration | Problems |
|---|---|---|
| Problem Solving (Basic) | 150 minutes | 2 |
| Problem Solving (Intermediate) | 150 minutes | 2 |

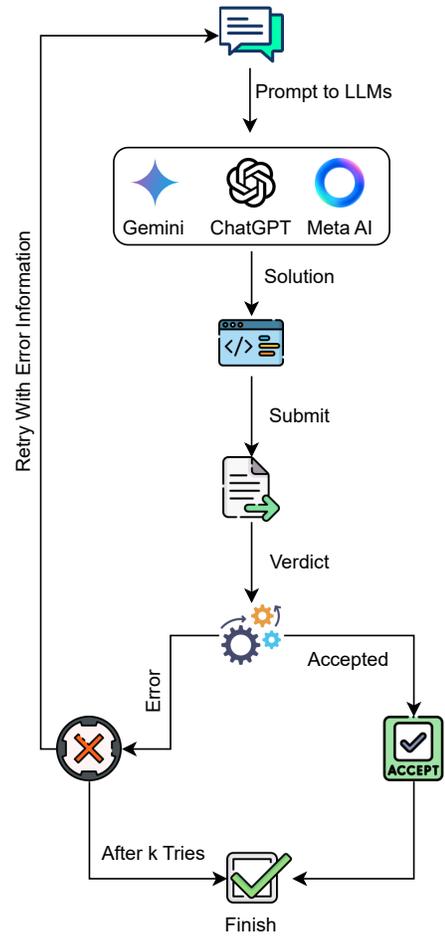

Figure 2: Workflow of solving problems

For both online and offline problem-solving, we directly copied the problem descriptions from the online judges and organized them according to our predefined prompts before inputting them into the LLMs. Once a solution, comprising both explanations and code was generated, we extracted our desired implementation. Furthermore, we post-processed error information to streamline the data before re-prompting the LLMs.

Concerning HackerRank certification tests, we faced limitations when attempting to copy the problem descriptions directly from the platform. To overcome this obstacle, we employed a third-party tool[13] for image-to-text conversion. This involved capturing the problem description image, converting it to text using the tool, and subsequently adapting it to fit our prompt structure.

## 5 Results

### 5.1 RQ$_1$: Success Rate

Out of the 98 problems selected from the LeetCode archive, ChatGPT successfully solved 70, accounting for 71.43% of the total problems. Meta AI's success rate was 58.16%, solving 57 problems, while Gemini solved 67, achieving a success rate of 68.37%.

For the 126 problems from Codeforces, ChatGPT, Meta AI, and Gemini solved 34, 21, and 10 problems, respectively, resulting in success rates of 26.98%, 16.67%, and 7.94%. ChatGPT outperformed both Gemini and Meta AI across both platforms, while Gemini and Meta AI showed varying performance depending on the platform.

The results of our experiment in online contests are shown in TABLE 3. We aimed to cover all contest types by participating in nine virtual contests on LeetCode and Codeforces. We have highlighted the best results achieved out of the three LLMs we used. Across 9 online contests, there were 49 problems in total, but the LLMs were only able to solve 8 of them. To break it down further, they solved only 2 out of 33 Codeforces problems and 6 out of 16 LeetCode problems.

[13]https://www.imagetotext.info/

**Table 3: Codeforces (Cf) and LeetCode (LC) virtual contests' results**

| Contests | Solved | Standing |
|---|---|---|
| Cf - Educational | 0 / 6 | |
| Cf - Division 4 | 1 / 7 | 22134 / 60343 |
| Cf - Division 3 | 1 / 8 | 16324 / 53980 |
| Cf - Division 2 | 0 / 6 | |
| Cf - Division 1 | 0 / 6 | |
| LC - Weekly 1 | 1 / 4 | 10383 / 22688 |
| LC - Biweekly 1 | 2 / 4 | 9999 / 25287 |
| LC - Weekly 2 | 1 / 4 | 15801 / 29736 |
| LC - Biweekly 2 | 2 / 4 | 12834 / 26241 |

We undertook the two certification tests on HackerRank as detailed in TABLE 2, utilizing ChatGPT, Gemini, and Meta AI. All three LLMs successfully obtained Problem-Solving certifications at the Basic and Intermediate levels. Additionally, these certificates have been included in the replication package.

> **Summary RQ$_1$:** On average, LLMs demonstrated higher success rates on LeetCode, with ChatGPT achieving a rate of 71.43%, and a moderate success rate on Codeforces at 26.98% (ChatGPT). Although LLMs excelled in HackerRank certifications, they encountered difficulties during virtual contests, particularly on Codeforces.

## 5.2 RQ$_2$: Performance Variation

Figure 3 illustrates the performance of various LLMs across 15 problem categories. It is evident that ChatGPT consistently achieved higher success rates in the first 11 categories. Conversely, Meta AI demonstrated superior performance in Number Theory (N) and Probability and Statistics (PS). On average, Gemini outperformed Meta AI. Notably, Meta AI encountered difficulties in six categories, where it registered the lowest success rates.

TABLE 4 illustrates how problems solved by difficulty and acceptance rate were distributed among the three LLMs. ChatGPT, Gemini, and Meta showed their strongest performance when the problems were easy and had a high acceptance rate. However, their performance noticeably declined as the difficulty of the problems increased to medium and hard, and the acceptance rate decreased to medium and low.

We conducted an experiment to see if the programming language used affects an LLM's problem-solving ability. We tackled 98 LeetCode problems from TABLE 1 using ChatGPT, using both C++ and Python separately. We followed identical procedures for both of the programming languages. The results showed that 70 problems were solved successfully using C++ and 67 with Python. We then conducted a chi-square [27] test of independence, with the null hypothesis being that programming language does not impact LLM performance. The resulting p-value was 0.76, which exceeds the significance level of 0.05. This means we failed to reject the null hypothesis, indicating that LLMs demonstrate no significant difference in performance when employing different programming languages.

> **Summary RQ$_2$:** LLMs display diverse performance trends. ChatGPT consistently succeeds across categories, Meta AI shines in Number Theory and Probability, and Gemini generally outperforms Meta AI. However, all LLMs struggle with harder problems and lower acceptance rates. Statistical tests confirmed that there is no noticeable difference in performance when employing different programming languages with LLMs for problem-solving.

## 5.3 RQ$_3$: Human Compatibility

LeetCode offers valuable insights into participants' accepted solutions, providing a comparative analysis of their performance regarding time and memory constraints. Figure 4 illustrates the percentage by which LLMs' performance exceeds that of users in terms of time efficiency and memory usage across the 98 problems detailed in TABLE 1.

Specifically, Gemini exhibited higher time efficiency, outperforming 65.79% of the users. Conversely, Meta AI showcased superior memory usage, surpassing 54.29% of the users, outperforming both Gemini and LeetCode. On average, LLMs outperformed 63.10% of users in time efficiency and 51.08% of users in memory usage.

However, unlike LeetCode, Codeforces does not offer comparable insights. Instead, it only provides information on the time taken to solve a problem and the memory it consumes.

When we compared the performance of LLMs with that of humans in the online contests presented in TABLE 3, which we participated in through LLMs, we observed varied results across different types of contests. We collected the standings of LLMs from the virtual contests they participated in and calculated the percentage by which LLMs outperformed other participants in those contests. LLMs performed moderately well in the weekly and biweekly contests hosted by LeetCode, where they outperformed an average of 53.7% of users. However, their performance was poor in the more challenging Codeforces contests (Division 1 and Division 2). They achieved better results in less difficult contests such as Division 4 (63.32%) and Division 3 (69.76%), but struggled in the Educational contest.

> **Summary RQ$_3$:** LLMs generally outperformed a significant portion of users, demonstrating strong performance in both time efficiency and memory usage when tested against problems from the LeetCode archive. However, in live contests, LLMs exhibited moderate performance in LeetCode contests but encountered challenges in harder Codeforces contests when compared to humans.

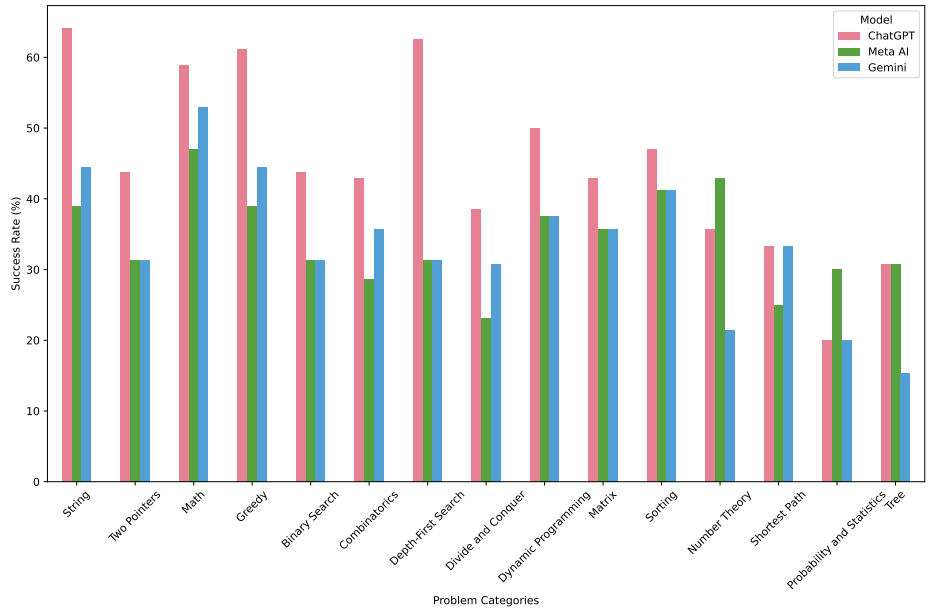

Figure 3: Success Rates Across Problem Categories

Table 4: Difficulty and Acceptance Range Wise Solved Problems by ChatGPT (C), Gemini (G), and Meta AI (M)

| Difficulty | Acceptance (C) | | | Acceptance (G) | | | Acceptance (M) | | |
|---|---|---|---|---|---|---|---|---|---|
| | High | Medium | Low | High | Medium | Low | High | Medium | Low |
| Easy | 23 | 17 | 2 | 18 | 16 | 3 | 22 | 17 | 3 |
| Medium | 18 | 18 | 5 | 8 | 12 | 7 | 10 | 14 | 2 |
| Hard | 8 | 11 | 1 | 3 | 8 | 2 | 4 | 5 | 1 |

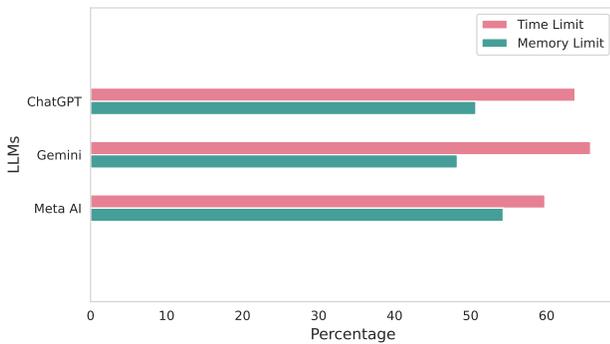

Figure 4: Comparison with users' performance with LLMs in terms of time and memory usage of LeetCode

## 6 Discussion
### 6.1 Action is Needed
Given the moderate to higher success rates of LLMs in solving diverse programming challenges (RQ$_1$) and their above-average performance compared to humans (RQ$_3$), it is imperative for online programming platforms to strategize how to address this trend. As LLMs continue to evolve rapidly with the advent of larger models and increased computational capabilities, it is foreseeable that they may emerge as leading performers on platforms such as Codeforces and LeetCode in the near future. If programming platforms do not take adequate measures to protect against LLM-related threats, they risk losing trust within the programming community.

HackerRank's Certification tests, recognized by the programming community [19], prevent users from copying problem descriptions to deter cheating. However, it is evident from our methodology (Subsection 4.4) that obtaining programming certificates listed in TABLE 2 can be accomplished quite easily. This exposes a vulnerability in the system, indicating insufficient protective measures against threats related to LLMs across different platforms.

Recent studies, such as the one by Idialu et al. [16], have demonstrated promising outcomes in distinguishing between code authored by humans and code generated by GPTs. Online programming platforms can adopt similar techniques to determine code authorship. However, their classifier is specifically designed for the challenges posed on CodeChef. Future research endeavors could explore the development of a multi-platform classifier with similar capabilities to counteract the influence of potent LLMs on various programming platforms.

## 6.2 Significance of Feedback

In our methodology, we utilized a chain-of-thought prompting approach [50] when our initial submission attempts (k=1) were not accepted by the online judges. This approach involved an iterative process where we fed the error details received from the judges back into the Language Model Models (LLMs) and made subsequent attempts up to a total of five times (k=5).

Following this approach, both ChatGPT and Gemini were able to resolve 12.24% of the 98 LeetCode problems, while 4.76% of the 126 Codeforces problems were resolved by both ChatGPT and Meta AI. The process of providing feedback from the online judge to the LLM served as a continuous learning process. It allowed the LLM to learn from its mistakes by understanding the specific errors encountered, which is consistent with the findings of Tong et al. [44] suggesting that LLMs learn from previous mistakes.

## 6.3 Challenges Faced by LLMs

The results of $RQ_1$ and $RQ_3$ show that all three LLMs encountered challenges in solving problems on Codeforces compared to LeetCode. To delve into the reasons, we analyzed the Division 4 contest from Codeforces and the Weekly 1 contest from LeetCode (see Table 3) to compare the average number of words per problem. We found that the Codeforces contest had an average of 274.14 words per problem, while the LeetCode contest had an average of 164.25 words per problem. This suggests that LLMs might struggle to understand larger problem descriptions, leading to poorer performance on Codeforces.

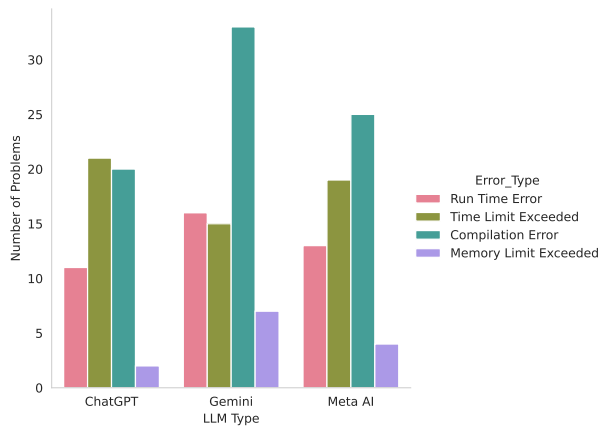

**Figure 5: Distribution of different errors across LLMs**

To gain deeper insights into the challenges encountered by LLMs during problem-solving, we provide an analysis of error types across different LLMs, as depicted in Figure 5. The distribution illustrates that Gemini and Meta AI encountered the most difficulties in generating compilable solutions for the problems. Conversely, ChatGPT struggled with producing accurate solutions within the specified time constraints of programming platforms. Despite these challenges, all three LLMs demonstrated considerable success in solving problems within the prescribed memory constraints.

Additionally, among the 224 problems assessed, 35 included images in their descriptions to aid programmers in better comprehending the problem. Despite not incorporating these images in the problem descriptions, ChatGPT successfully solved 60% of the problems, while Meta AI managed to solve 42.85% of them. Notably, LLMs demonstrated competence even when confronted with problems featuring images.

## 6.4 Number of Attempts

In our problem-solving methodology, we attempted each problem five times (k=5). We chose this number based on the belief that it represents an optimal balance for assessing the impact of providing feedback to LLMs. Given that our approach involves utilizing conversational LLMs, we aimed to avoid relying on paid APIs and instead manually solve problems through copying and pasting. Using higher values of k would have necessitated significant manual effort and time investment. Our decision aligns with previous research [9, 40], which also utilized lower values of k (<=3) in their studies. To verify our assumption, we conducted a pilot study with a higher k value of 10 during the Division 4 Codeforces contest. Interestingly, this adjustment did not alter the results presented in TABLE 3.

## 7 Implications

### 7.1 For Researchers

Our study's findings can be used to investigate LLMs' shortcomings in complex code generation. Our results reveal that LLMs face difficulties in generating solutions when problems in the hard and low acceptance range are given to them. Researchers associated with developing AI coding assistants like GitHub Copilot can incorporate our findings into their models to make improved coding assistant tools that will be able to tackle difficult problem-solving scenarios.

### 7.2 For Recruiters

Efforts [28] have been made to automatically gather candidates' online programming profile information to facilitate hiring. Our study findings indicate that LLMs perform moderately well on programming platforms and excel in obtaining online certificates related to programming skills. These insights provide recruiters with a means to verify the credibility of a candidate's online programming profile. However, while these platforms offer valuable information, recruiters should supplement their evaluation methods with interviews, practical assessments, or real-world projects. This holistic approach ensures a thorough understanding of candidates' skills, behaviors, and problem-solving abilities, enabling recruiters to make more informed hiring decisions.

### 7.3 For Practitioners

We conducted a study to assess the effectiveness of various online programming platforms using multiple LLMs across various programming challenges. This study will help developers associated with these platforms identify any potential vulnerabilities in their systems and work towards making their platforms more secure against LLMs. Additionally, our findings will enable online judge problem-setters to create programming challenges that are more challenging for LLMs to solve.

## 8 Threats to Validity

Although we have diligently strived for accuracy, there is a chance that validity threats could influence the study's results. Utilizing Runeson et al.'s classification [39], we scrutinize multiple factors that might jeopardize the study's integrity.

### 8.1 Construct Validity

Construct validity concerns whether our chosen operational measures effectively address the main inquiries in our research. In our case, the design of prompts used for solving programming problems on various online judges might potentially limit LLMs' ability to understand programs. This limitation arises from the nature and structure of the prompts, which may not fully represent the complexity or diversity of real-world problem-solving scenarios. To address this concern, we followed prior research and adhered to OpenAI's guidelines for prompt design (Subsection 4.4).

To report the results of our research questions, we used various quantitative metrics such as success rate, acceptance rate, time efficiency, and memory efficiency. These metrics are well-established and verified by online judges. While not all platforms provide every metric, this did not affect our study's results, as most quantitative metrics are commonly available across online judges.

### 8.2 Internal Validity

In this study, we focused exclusively on conversational LLMs. Instead of utilizing the more advanced paid versions of ChatGPT and Gemini, such as GPT-4 based ChatGPT Plus[14] and Gemini Advanced[15], we opted for their free versions. We deliberately chose to exclusively use open-source free conversational LLMs, ensuring the study's generalizability, given that Meta AI does not offer any paid versions. However, this technological decision and assumption may not fully represent the capabilities of LLMs, potentially impacting the study's generalizability and the accuracy of the conclusions drawn from the experiment. To address this concern, we can refer to the work of Hans et al. [13], where they compared the coding capabilities of GPT-3.5 with GPT-4 on LeetCode problems and found negligible differences in their performance.

### 8.3 External Validity

Concerning external validity threats, the concern revolves around the generalizability of findings derived from LLMs' problem-solving behavior. The non-deterministic nature [34] of LLMs, where its solutions may vary for the same problem, poses a challenge to the reliability and generalizability of results. Additionally, selecting problems solely from LeetCode and Codeforces might limit the representation of the entire spectrum of programming challenges, potentially impacting the broader applicability of the study's findings beyond these specific platforms. This limitation could affect the extent to which conclusions drawn from LLMs' performance on these platforms can be extrapolated to other problem sets or programming environments.

### 8.4 Conclusion Validity

Regarding conclusion validity threats, the concerns center around the connection between outcomes and the treatments administered in the study. The dataset's restricted number of problems raises a potential concern about the solidity of statistical conclusions. This limitation could impede the generalizability of findings because a smaller dataset might not sufficiently reflect the wider problem landscape. To address this, we have curated the problems based on diverse problem categories, difficulty levels, and acceptance range levels to compile a more comprehensive dataset.

## 9 Conclusion and Future Work

This study aimed to evaluate the impact of LLMs on Competitive Programming across diverse platforms through a comprehensive analysis. Our investigation unveiled a nuanced spectrum of results across 15 categories, varying difficulty levels, and acceptance ranges. Notably, LLMs demonstrated high accuracy on LeetCode (71.43%) yet performed less optimally on Codeforces. While excelling in tests on platforms like HackerRank, their performance in live Codeforces contests was less favorable. Moreover, We conclude this study by highlighting the need for the programming community to devise measures to address the potential risks posed by LLMs to the established norms of programming platforms and advise recruiters to exercise caution when evaluating candidates.

For future exploration, efforts should aim to expand the range of platforms and problem sets analyzed while incorporating a broader selection of LLMs.

## Acknowledgments

This work was supported in part by the Natural Sciences and Engineering Research Council of Canada (NSERC) Discovery grants, the John R. Evans Leaders Fund (JELF) of the Canada Foundation for Innovation (CFI), and the industry-stream NSERC CREATE graduate program on Software Analytics Research (SOAR) grants.

## References

[1] Imtiaz Ahmed, Mashrafi Kajol, Uzma Hasan, Partha Protim Datta, Ayon Roy, and Md Rokonuzzaman Reza. 2023. Chatgpt vs. bard: A comparative study. *UMBC Student Collection* (2023).

[2] Ajmain I. Alam, Palash R. Roy, Farouq Al-Omari, Chanchal K. Roy, Banani Roy, and Kevin A. Schneider. 2023. GPTCloneBench: A comprehensive benchmark of semantic clones and cross-language clones using GPT-3 model and Semantic-CloneBench. In *2023 IEEE International Conference on Software Maintenance and Evolution (ICSME)*. 1–13. https://doi.org/10.1109/ICSME58846.2023.00013

[3] Ian Nery Bandeira, Thiago Veras Machado, Vitor F Dullens, and Edna Dias Canedo. 2019. Competitive programming: A teaching methodology analysis applied to first-year programming classes. In *2019 IEEE Frontiers in Education Conference (FIE)*. IEEE, 1–8.

[4] Ali Borji and Mehrdad Mohammadian. 2023. Battle of the Wordsmiths: Comparing ChatGPT, GPT-4, Claude, and Bard. *GPT-4, Claude, and Bard (June 12, 2023)* (2023).

[5] Victor R Basili1 Gianluigi Caldiera and H Dieter Rombach. 1994. The goal question metric approach. *Encyclopedia of software engineering* (1994), 528–532.

[6] Mark Chen, Jerry Tworek, Heewoo Jun, Qiming Yuan, Henrique Ponde de Oliveira Pinto, Jared Kaplan, Harri Edwards, Yuri Burda, Nicholas Joseph, Greg Brockman, et al. 2021. Evaluating large language models trained on code. *arXiv preprint arXiv:2107.03374* (2021).

[7] Tristan Coignion, Clément Quinton, and Romain Rouvoy. 2024. A Performance Study of LLM-Generated Code on C2. In *EASE'24 - 28th International Conference*

---

[14] https://openai.com/index/chatgpt-plus
[15] https://gemini.google.com/advanced

on Evaluation and Assessment in Software Engineering (Proceedings of the 28th International Conference on Evaluation and Assessment in Software Engineering (EASE'24)). Salerno, Italy. https://hal.science/hal-04525620

[8] Arghavan Moradi Dakhel, Vahid Majdinasab, Amin Nikanjam, Foutse Khomh, Michel C Desmarais, and Zhen Ming Jack Jiang. 2023. Github copilot ai pair programmer: Asset or liability? *Journal of Systems and Software* 203 (2023), 111734.

[9] Hampus Ekedahl and Vilma Helander. 2023. Can artificial intelligence replace humans in programming?

[10] Yunhe Feng, Sreecharan Vanam, Manasa Cherukupally, Weijian Zheng, Meikang Qiu, and Haihua Chen. 2023. Investigating Code Generation Performance of Chat-GPT with Crowdsourcing Social Data. In *Proceedings of the 47th IEEE Computer Software and Applications Conference*. 1–10.

[11] James Finnie-Ansley, Paul Denny, Brett A Becker, Andrew Luxton-Reilly, and James Prather. 2022. The robots are coming: Exploring the implications of openai codex on introductory programming. In *Proceedings of the 24th Australasian Computing Education Conference*. 10–19.

[12] Daniel Fried, Armen Aghajanyan, Jessy Lin, Sida Wang, Eric Wallace, Freda Shi, Ruiqi Zhong, Wen-tau Yih, Luke Zettlemoyer, and Mike Lewis. 2022. Incoder: A generative model for code infilling and synthesis. *arXiv preprint arXiv:2204.05999* (2022).

[13] Felix HANS. [n. d.]. GPT-3.5/4-Is the programming performance declining over time? *focus* 7, 8 ([n. d.]), 9. https://doi.org/10.13140/RG.2.2.30950.80966

[14] Felix Hans. 2023. ChatGPT vs. Bard - Which is better at solving coding problems? *10.13140/RG.2.2.36200.65284* (08 2023).

[15] Jocelyn Harper. 2022. Interview insight: How to get the job. In *A Software Engineer's Guide to Seniority: A Guide to Technical Leadership*. Springer, 19–28.

[16] Oseremen Joy Idialu, Noble Saji Mathews, Rungroj Maipradit, Joanne M Atlee, and Mei Nagappan. 2024. Whodunit: Classifying Code as Human Authored or GPT-4 Generated–A case study on CodeChef problems. *arXiv preprint arXiv:2403.04013* (2024).

[17] Baskhad Idrisov and Tim Schlippe. 2024. Program Code Generation with Generative AIs. *Algorithms* 17, 2 (2024), 62.

[18] Sandeep Kaur Kuttal, Xiaofan Chen, Zhendong Wang, Sogol Balali, and Anita Sarma. 2021. Visual Resume: Exploring developers' online contributions for hiring. *Information and Software Technology* 138 (2021), 106633.

[19] Paola A Leon Alarcon. 2024. *Understanding the Relevance, Efficiency and Efficacy of Timed Coding Assessments in the Software Engineering Industry*. Ph.D. Dissertation. Massachusetts Institute of Technology.

[20] Sila Lertbanjongngam, Bodin Chinthanet, Takashi Ishio, Raula Gaikovina Kula, Pattara Leelaprute, Bundit Manaskasemsak, Arnon Rungsawang, and Kenichi Matsumoto. 2022. An empirical evaluation of competitive programming ai: A case study of alphacode. In *2022 IEEE 16th International Workshop on Software Clones (IWSC)*. IEEE, 10–15.

[21] Raymond Li, Loubna Ben Allal, Yangtian Zi, Niklas Muennighoff, Denis Kocetkov, Chenghao Mou, Marc Marone, Christopher Akiki, Jia Li, Jenny Chim, et al. 2023. Starcoder: may the source be with you! *arXiv preprint arXiv:2305.06161* (2023).

[22] Yujia Li, David Choi, Junyoung Chung, Nate Kushman, Julian Schrittwieser, Rémi Leblond, Tom Eccles, James Keeling, Felix Gimeno, Agustin Dal Lago, et al. 2022. Competition-level code generation with alphacode. *Science* 378, 6624 (2022), 1092–1097.

[23] Jiawei Liu, Chunqiu Steven Xia, Yuyao Wang, and Lingming Zhang. 2024. Is your code generated by chatgpt really correct? rigorous evaluation of large language models for code generation. *Advances in Neural Information Processing Systems* 36 (2024).

[24] Zhijie Liu, Yutian Tang, Xiapu Luo, Yuming Zhou, and Liang Feng Zhang. 2024. No need to lift a finger anymore? Assessing the quality of code generation by ChatGPT. *IEEE Transactions on Software Engineering* (2024).

[25] Ziyang Luo, Can Xu, Pu Zhao, Qingfeng Sun, Xiubo Geng, Wenxiang Hu, Chongyang Tao, Jing Ma, Qingwei Lin, and Daxin Jiang. 2023. Wizardcoder: Empowering code large language models with evol-instruct. *arXiv preprint arXiv:2306.08568* (2023).

[26] Aman Madaan, Niket Tandon, Prakhar Gupta, Skyler Hallinan, Luyu Gao, Sarah Wiegreffe, Uri Alon, Nouha Dziri, Shrimai Prabhumoye, Yiming Yang, et al. 2024. Self-refine: Iterative refinement with self-feedback. *Advances in Neural Information Processing Systems* 36 (2024).

[27] Mary L McHugh. 2013. The chi-square test of independence. *Biochemia medica* 23, 2 (2013), 143–149.

[28] Saurav Muke, Samruddhi Ahire, Geetanjali Kale, and Pranali Navghare. 2024. ICode-An Unified Competitive Coding Profile Platform. In *2024 International Conference on Emerging Smart Computing and Informatics (ESCI)*. IEEE, 1–5.

[29] Nathalia Nascimento, Paulo Alencar, and Donald Cowan. 2023. Artificial Intelligence Versus Software Engineers: An Evidence-based Assessment Focusing on Non-functional Requirements. (2023).

[30] Nathalia Nascimento, Paulo Alencar, and Donald Cowan. 2023. Comparing software developers with chatgpt: An empirical investigation. *arXiv preprint arXiv:2305.11837* (2023).

[31] Nhan Nguyen and Sarah Nadi. 2022. An empirical evaluation of GitHub copilot's code suggestions. In *Proceedings of the 19th International Conference on Mining Software Repositories*. 1–5.

[32] Erik Nijkamp, Bo Pang, Hiroaki Hayashi, Lifu Tu, Huan Wang, Yingbo Zhou, Silvio Savarese, and Caiming Xiong. 2022. Codegen: An open large language model for code with multi-turn program synthesis. *arXiv preprint arXiv:2203.13474* (2022).

[33] N Nikolaidis, K Flamos, D Feitosa, A Chatzigeorgiou, and A Ampatzoglou. 2023. The End of an Era: Can Ai Subsume Software Developers. *Evaluating Chatgpt and Copilot Capabilities Using Leetcode Problems* (2023).

[34] Shuyin Ouyang, Jie M Zhang, Mark Harman, and Meng Wang. 2023. LLM is Like a Box of Chocolates: the Non-determinism of ChatGPT in Code Generation. *arXiv preprint arXiv:2308.02828* (2023).

[35] Harikumar Pallathadka, V Hari Leela, Sushant Patil, BH Rashmi, Vipin Jain, and Samrat Ray. 2022. Attrition in software companies: Reason and measures. *Materials Today: Proceedings* 51 (2022), 528–531.

[36] Dipendra Pant, Dhiraj Pokhrel, and Prakash Poudyal. 2022. Automatic Software Engineering Position Resume Screening using Natural Language Processing, Word Matching, Character Positioning, and Regex. In *2022 5th International Conference on Advanced Systems and Emergent Technologies (IC_ASET)*. IEEE, 44–48.

[37] Palash R. Roy, Ajmain I. Alam, Farouq Al-omari, Banani Roy, Chanchal K. Roy, and Kevin A. Schneider. 2023. Unveiling the Potential of Large Language Models in Generating Semantic and Cross-Language Clones. In *2023 IEEE 17th International Workshop on Software Clones (IWSC)*. 22–28. https://doi.org/10.1109/IWSC60764.2023.00011

[38] Palash Ranjan Roy, Md. Noushin Islam, Labiba Tasfiya Jeba, Md. Adnanul Haq, Iffat Afsara Prome, Mohammad Kaykobad, and Tanvir Kaykobad. 2022. A Study on Paper and Author Ranking. In *2022 International Conference on Innovations in Science, Engineering and Technology (ICISET)*. 545–549. https://doi.org/10.1109/ICISET54810.2022.9775821

[39] Per Runeson and Martin Höst. 2009. Guidelines for conducting and reporting case study research in software engineering. *Empirical software engineering* 14 (2009), 131–164.

[40] Fardin Ahsan Sakib, Saadat Hasan Khan, and AHM Karim. 2023. Extending the frontier of chatgpt: Code generation and debugging. *arXiv preprint arXiv:2307.08260* (2023).

[41] Shashi Kant Singh, Shubham Kumar, and Pawan Singh Mehra. 2023. Chat GPT & Google Bard AI: A Review. In *2023 International Conference on IoT, Communication and Automation Technology (ICICAT)*. IEEE, 1–6.

[42] Shuo Sun, Yuchen Zhang, Jiahuan Yan, Yuze Gao, Donovan Ong, Bin Chen, and Jian Su. 2023. Battle of the Large Language Models: Dolly vs LLaMA vs Vicuna vs Guanaco vs Bard vs ChatGPT–A Text-to-SQL Parsing Comparison. *arXiv preprint arXiv:2310.10190* (2023).

[43] Gemini Team, Rohan Anil, Sebastian Borgeaud, Yonghui Wu, Jean-Baptiste Alayrac, Jiahui Yu, Radu Soricut, Johan Schalkwyk, Andrew M Dai, Anja Hauth, et al. 2023. Gemini: a family of highly capable multimodal models. *arXiv preprint arXiv:2312.11805* (2023).

[44] Yongqi Tong, Dawei Li, Sizhe Wang, Yujia Wang, Fei Teng, and Jingbo Shang. 2024. Can LLMs Learn from Previous Mistakes? Investigating LLMs' Errors to Boost for Reasoning. *arXiv preprint arXiv:2403.20046* (2024).

[45] Hugo Touvron, Thibaut Lavril, Gautier Izacard, Xavier Martinet, Marie-Anne Lachaux, Timothée Lacroix, Baptiste Rozière, Naman Goyal, Eric Hambro, Faisal Azhar, et al. 2023. Llama: Open and efficient foundation language models. *arXiv preprint arXiv:2302.13971* (2023).

[46] Priyan Vaithilingam, Tianyi Zhang, and Elena L Glassman. 2022. Expectation vs. experience: Evaluating the usability of code generation tools powered by large language models. In *Chi conference on human factors in computing systems extended abstracts*. 1–7.

[47] Yue Wang, Hung Le, Akhilesh Deepak Gotmare, Nghi DQ Bui, Junnan Li, and Steven CH Hoi. 2023. Codet5+: Open code large language models for code understanding and generation. *arXiv preprint arXiv:2305.07922* (2023).

[48] Yue Wang, Weishi Wang, Shafiq Joty, and Steven CH Hoi. 2021. Codet5: Identifier-aware unified pre-trained encoder-decoder models for code understanding and generation. *arXiv preprint arXiv:2109.00859* (2021).

[49] Szymon Wasik, Maciej Antczak, Jan Badura, Artur Laskowski, and Tomasz Sternal. 2018. A survey on online judge systems and their applications. *ACM Computing Surveys (CSUR)* 51, 1 (2018), 1–34.

[50] Jason Wei, Xuezhi Wang, Dale Schuurmans, Maarten Bosma, Fei Xia, Ed Chi, Quoc V Le, Denny Zhou, et al. 2022. Chain-of-thought prompting elicits reasoning in large language models. *Advances in neural information processing systems* 35 (2022), 24824–24837.

[51] Jules White, Quchen Fu, Sam Hays, Michael Sandborn, Carlos Olea, Henry Gilbert, Ashraf Elnashar, Jesse Spencer-Smith, and Douglas C Schmidt. 2023. A prompt pattern catalog to enhance prompt engineering with chatgpt. *arXiv preprint arXiv:2302.11382* (2023).